\documentclass[pra,twocolumn,preprintnumbers,amsmath,amssymb,superscriptaddress,longbibliography,aps]{revtex4-2}

\usepackage{color}
\usepackage{graphicx}
\usepackage{dcolumn}
\usepackage{bm}
\usepackage{hyperref}
\usepackage{enumitem}
\usepackage{balance}
\usepackage{comment}
\usepackage{cleveref}

\usepackage{amsthm}

\begin{document}
\title{Shortcut to adiabaticity improvement of STIRAP based qubit rotation}

\author{Khayla Black}
\affiliation{New York University Shanghai, NYU-ECNU Institute of Physics at NYU Shanghai, Shanghai Frontiers Science Center of Artificial Intelligence and Deep Learning, 567 West Yangsi Road, Shanghai, 200126, China.}

\author{Xi Chen}
\affiliation{Instituto de Ciencia de Materiales de Madrid (CSIC), Cantoblanco, E-28049 Madrid, Spain}

\author{Tim Byrnes}
 \email{tim.byrnes@nyu.edu}
\affiliation{New York University Shanghai, NYU-ECNU Institute of Physics at NYU Shanghai, Shanghai Frontiers Science Center of Artificial Intelligence and Deep Learning, 567 West Yangsi Road, Shanghai, 200126, China.}
\affiliation{State Key Laboratory of Precision Spectroscopy, School of Physical and Material Sciences, East China Normal University, Shanghai 200062, China}
\affiliation{Center for Quantum and Topological Systems (CQTS), NYUAD Research Institute, New York University Abu Dhabi, UAE.}
\affiliation{Department of Physics, New York University, New York, NY 10003, USA}

\begin{abstract}
Robust quantum control is essential for the development of quantum computers, which rely on precise manipulation of qubits. One form of quantum control is stimulated Raman adiabatic passage (STIRAP), which ordinarily is a state transfer protocol but was extended by Kis and Renzoni  (Phys. Rev. A 65, 032318 (2002)) to perform qubit rotations. Shortcut methods to adiabaticity for STIRAP have been shown to speed up adiabatic processes, beyond the adiabatic criterion, with high fidelity. Here, we apply shortcut to adiabaticity methods to the STIRAP qubit rotation scheme to improve the performance of quantum logic gates. The scheme can be implemented via direct connections between ground states in a 4-level $\Lambda$ system or effective connections in a 5-level $\Lambda$ system with modified pulses that implement transitionless quantum driving via the addition of a counterdiabatic driving term. We show that the extended shortcut to adiabaticity method serves to improve the fidelity of qubit rotations in the diabatic regime. 
\end{abstract}

\date{\today}

\maketitle

\section{Introduction \label{i: intro}}

With recent developments in quantum technologies, improving quantum control is increasingly a necessity on many experimental platforms. Quantum computing requires high-fidelity control of quantum systems, but this is a challenging task requiring optimization of numerous parameters. Stimulated Raman adiabatic passage (STIRAP) is a protocol introduced by Gaubatz, Bergmann and co-workers that has been widely used to realize robust quantum control over a variety of quantum systems \cite{gaubatz_population_1990, VITANOV_stirap_coherent, bergman_coherent, kuklinski_adiabatic, Vitanov2017, bergmann2015perspective}. STIRAP makes use of dark states to create an effective coupling between ground states. The parameters of the Hamiltonian are then adjusted such that the state adiabatically evolves from one ground state to the other, in the case of an archetypal 3-level $\Lambda$ system. Occupation of the excited state is precisely zero in the ideal case, and emission from the intermediate state is highly suppressed by the use of dark states, which are eigenstates of the Hamiltonian that have exactly zero amplitude of the excited state \cite{Vitanov2017}. The zero transition amplitudes of the dark states prevent decoherence via spontaneous emission of intermediate states, allowing robust state transfer \cite{Vitanov2017,byrnes2021quantum}. 

Adiabatic passage techniques have been applied to numerous contexts including ultracold molecular formation \cite{Winkler2007, Danzl2008, danzl_ultracold_2010}, coupled acoustics cavities \cite{Shen2019}, quantum information processing \cite{Weidt2016, Webster2013,thomasen2016ultrafast}, Rydberg excitations \cite{PhysRevLett.100.170504,rao2014robust,khazali2020fast,idlas2016entanglement,saffman2010quantum}, atomic ensembles \cite{beterov2013quantum,beterov2020application,hussain2014geometric,ortiz2018adiabatic}, and superconducting quantum circuits \cite{kumar_stimulated_2016, xu_coherent_2016,siewert2004applications}. Conventionally, STIRAP is a state transfer protocol, where there is a population transfer achieved from a known initial state to a final state. It can however also be adapted into a quantum gate, where the net result is a unitary rotation \cite{Kis2002,lacour2006arbitrary,rousseaux2013arbitrary,kral2007colloquium}.  Kis and Renzoni introduced a method to perform STIRAP-based qubit rotations (henceforth called SQR) using two consecutive STIRAP sequences  \cite{Kis2002}. The scheme operates on a 4-level $\Lambda$ system (Fig. \ref{fig1}(a)) where the qubit is defined as a linear combination of two of the three ground states. The first STIRAP pulse transfers the state to a linear combination of all three ground states.  The second STIRAP pulse then returns the state to the logical states in such a way that the final state is modified according to the desired unitary rotation. 
The scheme has been shown to have robust ability in implementing quantum logic gates \cite{Kis2002}. SQR has been experimentally verified in a trapped $^{40}\text{Ca}^+$ ion \cite{realization_toyoda}.

One of the limitations of such adiabatic passage techniques is that it must work in the adiabatic regime, such that the Hamiltonian is modified sufficiently slowly. Faster changes produces diabatic excitations, which can be a cause of errors in the protocol. To overcome these limitations, so-called shortcut to adiabaticity methods have been developed, where effective adiabadicity can be achieved while working in a diabatic regime \cite{chen_phys_nodate-2,unanyan1997laser,couvert2008optimal,chen2010transient,torrontegui2013shortcuts,del2013shortcuts,guery2019shortcuts, Hegade2021,del2019focus, Vepsalainen2019}. One method of performing shortcut to adiabaticity is via counter-diabatic driving proposed by Demirplak and Rice \cite{rice_adiabatic_2003}, mathematically formulated as transitionless quantum driving by Berry \cite{berry_transitionless_2009}. Transitionless quantum driving involves the addition of a counterdiabatic driving term to an interacting Hamiltonian. The addition of the counterdiabatic term to a slowly varying time-dependent Hamiltonian creates a Hamiltonian that drives the system exactly, preventing any transitions between eigenstates. This improves the performance of time evolution and the fidelity of time evolved states. This method has been applied to increase STIRAP efficiency in a standard 3-level $\Lambda$ system by Du, Zhu, and co-workers. \cite{du_2016}. The counterdiabatic term is added to the canonical 3-level $\Lambda$ system STIRAP configuration via modified Rabi pulses that implement both the original STIRAP pulses and the counterdiabatic driving. The enhanced performance of STIRAP was numerically and experimentally verified \cite{du_2016}.

In this paper, we provide a method for performing STIRAP-based quantum logic gates with improved performance using shortcut to adiabaticity methods. Our starting point is the SQR Hamiltonian for the qubit rotation of Ref. \cite{Kis2002}. We then use the transitionless quantum driving algorithm to obtain the counterdiabatic driving term. We implement the counterdiabatic driving in two ways, first by directly adding the associated terms as given by Berry's scheme \cite{berry_transitionless_2009}, by adding direct connections between the ground states (Fig. \ref{fig1}(a)). We also implement the scheme via modified Rabi pulses creating effective connections between three ground states and two excited states in a 5-level $\Lambda$ system (Fig. \ref{fig1}(b)), in an approach similar to Ref. \cite{du_2016}. We support our theoretical results with numerical simulations showing a fidelity improvement in the shortcut to adiabaticity based quantum logic gates in the two  implementations.

\section{Theory of STIRAP-based qubit rotation}
\label{sec:theory}

\subsection{Standard theory without counterdiabatic driving}
\label{ii a: SQR theory}

\begin{figure}[t]
\includegraphics[width=\linewidth]{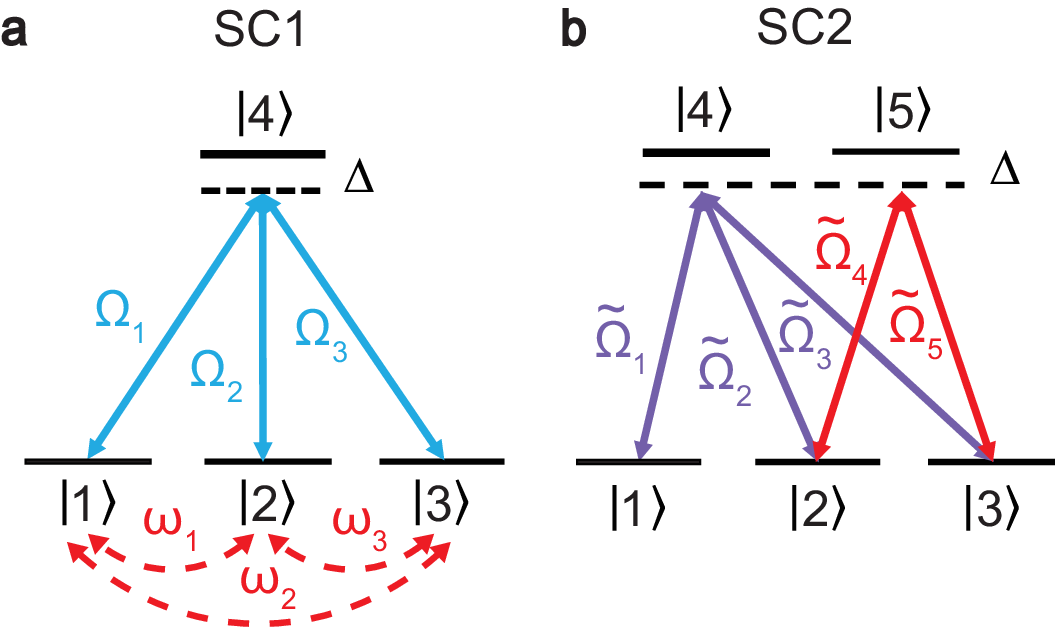}
\caption{Schemes for shortcut to adiabaticity improvement of SQR. a) The SQR scheme with direct ground state counterdiabatic driving (scheme SC1).
Here $ \Omega_i $ with $i \in \{1,2,3 \} $ denote SQR transitions as given in (\ref{kisrenzoniham}), and $ \omega_i$ are the counterdiabatic driving transitions defined in (\ref{omegaham}).  The qubit is defined by the ground states $\vert 1 \rangle$ and $\vert 2 \rangle$.  b) The SQR scheme as given with effective counterdiabatic driving via Raman transitions (scheme SC2).   The qubit is defined by the ground states $\vert 1 \rangle$ and $\vert 2 \rangle$; $ \vert 3 \rangle $ is the same auxiliary level as that required in the original SQR scheme.   $\vert 4 \rangle$ and $\vert 5 \rangle$ are two excited states that are required to implement the STIRAP pulses as well as the counterdiabatic driving.  $\tilde{\Omega}_1$, $\tilde{\Omega}_2$, $\tilde{\Omega}_3$ involve the original SQR pulses together modified by the Raman counterdiabatic driving. $\tilde{\Omega}_4$, and $\tilde{\Omega}_5$ additional corrections such as to implement the counterdiabatic driving required between states $\vert 2 \rangle$ and $\vert 3 \rangle$ by a Raman transition.  
\label{fig1}  }
\end{figure}

Our starting point is the SQR Hamiltonian introduced in Ref. \cite{Kis2002}. The energy levels in the system are shown in Fig. \ref{fig1}(a). The system contains three ground states $\vert 1 \rangle$, $\vert 2 \rangle$ and $\vert 3 \rangle$ where the qubit is defined by states $\vert 1 \rangle$ and $\vert 2 \rangle$. Our initial state is a linear combination of these two ground states 

\begin{equation}
    \vert \psi_0 \rangle = \alpha \vert 1 \rangle + \beta \vert 2 \rangle.
\end{equation}
The third ground state $ | 3 \rangle $ serves as an auxiliary state which will
be occupied only in the intermediate phase of the rotation procedure. All states are coupled to an excited state $\vert 4 \rangle$
by different laser fields shown as solid lines in Fig. \ref{fig1}(a). 

\begin{figure}[t]
\includegraphics[width=\linewidth]{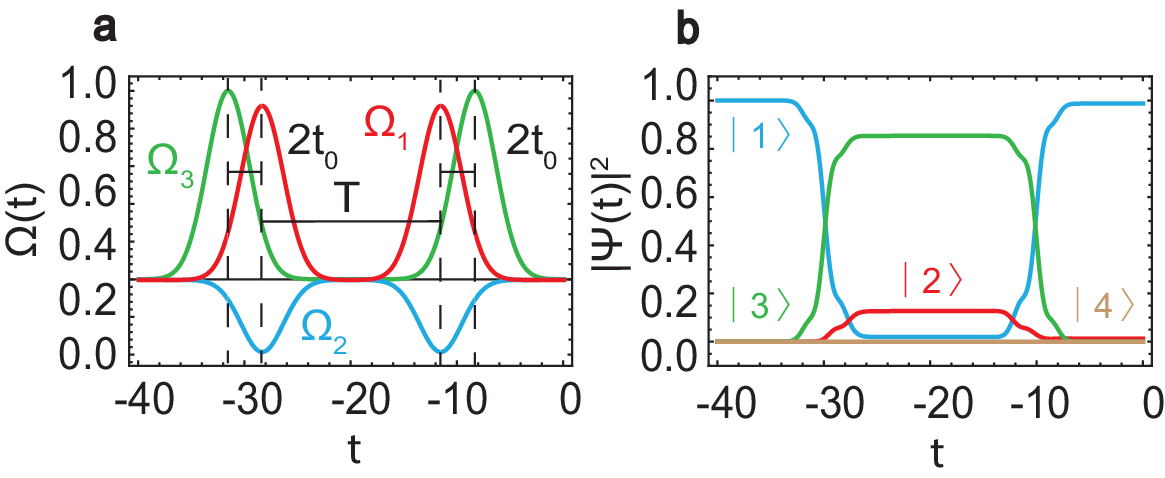}
\caption{The standard SQR pulse scheme and evolution. (a) Pulses driving the SQR scheme given by (\ref{pulse1}), (\ref{pulse2}), and (\ref{pulse3}). The width and separation of these pulses is defined by $T$, $t_0$, and $\sigma$ as marked.  (b) The time evolution of the Hamiltonian (\ref{kisrenzoniham}) driven by the pulses in (a) starting from an initial state $ \vert \psi_0 \rangle = \vert 1 \rangle $.  The parameters of this rotation are $\chi = \frac{\pi}{8}$, $\eta = \pi$, and $\delta = 0$, corresponding to an identity gate.
\label{fig2}  }
\end{figure}

The Hamiltonian for this system in the rotating frame is given by
\begin{equation}
    H_{\text{SQR}} = \hbar \Delta \vert 4 \rangle \langle 4 \vert + \frac{\hbar}{2} \sum^3_{i = 1}(\Omega_i \vert i \rangle \langle 4 \vert + \text{H.c.}), \label{kisrenzoniham}
\end{equation} 
where H.c. is the Hermitian conjugate. The SQR procedure consists of two STIRAP sequences. The pulses are applied in the usual counter-intuitive order, i.e. the Stokes pulse arrives before
the pump pulse. Pulses $\Omega_1(t)$ and $\Omega_2(t)$ are the pump pulses, and the pulse $\Omega_3(t)$ is the Stokes pulse. The pulses are defined according to 
\begin{align}
  \Omega_1(t) & = \Omega_0
   \text{cos}(\chi) \biggl( e^{\frac{-(t + T/2 + t_0)^2)}{2 \sigma^2}} + e^{\frac{-(t + T/2 - t_0)^2)}{2 \sigma^2}} \biggr) \label{pulse1} \\
    \Omega_2(t) & = \Omega_0
   \text{sin}(\chi) e^{i\eta} \biggl(e^{\frac{-(t + T/2 + t_0)^2)}{2 \sigma^2}} + e^{\frac{-(t + T/2 - t_0)^2)}{2 \sigma^2}} \biggr) \label{pulse2} \\
     \Omega_3(t) & = \Omega_0
   e^{i\delta} \biggl( e^{\frac{-(t + T/2 - t_0)^2)}{2 \sigma^2}} + e^{\frac{-(t + T/2 + t_0)^2)}{2 \sigma^2}} \biggr) .\label{pulse3}    
\end{align}
The form of these pulses is depicted in Fig. \ref{fig2}(a). 

The pump pulses $\Omega_1(t)$ and $\Omega_2(t)$ define a dark state 
\begin{equation}
    \vert S_D \rangle = -\text{sin} \chi \vert 1 \rangle + e^{i\eta}\text{cos}\chi \vert 2 \rangle.
\end{equation} 
We define a state orthogonal to this dark state known as the bright state 
\begin{equation}
    \vert S_B \rangle = \text{cos} \chi \vert 1 \rangle + e^{i\eta}\text{sin}\chi \vert 2 \rangle.
\end{equation} 
We can rewrite our initial qubit state into a superposition of $\vert S_D \rangle$ and $\vert S_B \rangle$ as 
\begin{equation}
    \vert \psi_0 \rangle = \langle S_D \vert \psi_0 \rangle \vert S_D \rangle + \langle S_B \vert \psi_0 \rangle \vert S_B \rangle ,\label{initqubitstate} 
\end{equation} 
where
\begin{align}
    \langle S_D \vert \psi_0 \rangle & = -\alpha \text{sin}\chi + \beta e^{-i\eta}\text{cos}\chi \\
     \langle S_B \vert \psi_0 \rangle & = \alpha \text{cos}\chi + \beta e^{-i\eta}\text{sin}\chi.   
\end{align}

Solving the Schrodinger equation in the adiabatic limit leaves the system in a superposition of the three ground states according to \
\begin{equation}
    \vert \psi_{\text{int}} \rangle = \langle S_D \vert \psi_0 \rangle \vert S_D \rangle - \langle S_B \vert \psi_0 \rangle \vert 3 \rangle.
    \label{intermediatestate}
\end{equation}
Comparing this with the initial qubit state (\ref{initqubitstate}), we see that the dark component of the state remains untouched, but the bright component has been transferred to $\vert 3 \rangle$. The second STIRAP pulse then maps this component back to the subspace of the qubit states ${\vert 1 \rangle, \vert 2 \rangle}$ according to 
\begin{equation}
    \vert 3 \rangle \rightarrow e^{-i\delta} \vert S_B \rangle. \label{rotation}
\end{equation} 
Applying this to (\ref{intermediatestate}), the final state is equal to 
\begin{equation}
    \vert \psi_{\text{f}} \rangle = \langle S_D \vert \psi_0 \rangle \vert S_D \rangle + e^{-i\delta} \langle S_B \vert \psi_0 \rangle \vert S_B \rangle.
\end{equation} 

Comparing this result to the initial state, we may write this as a unitary rotation by an angle $\delta$ 
\begin{equation}
    \vert \psi_{\text{f}} \rangle = e^{-i \frac{\delta}{2}} U_{\hat{n}}(\delta)\vert \psi_0 \rangle ,\label{rotationformula}
\end{equation} 
with axis $n = (\text{sin}2\chi\text{cos}\eta, \text{sin}2\chi\text{sin}\eta, \text{cos} 2\eta)$, where
\begin{equation}
    U_{\hat{n}}(\zeta) = \text{exp}(-i\frac{\zeta}{2}\hat{n}\cdot \hat{\sigma}) = \text{cos}\frac{\zeta}{2} - i \hat{n} \cdot \hat{\sigma} \text{sin}\frac{\zeta}{2}.
\end{equation}
This demonstrates the fact that these two STIRAP sequences collectively form a rotation of the qubit.

\subsection{Including counterdiabatic driving} 
\label{ii b: SQR with cd driving theory}

Our aim is to improve the efficiency of SQR by using shortcut to adaibaticity methods. To achieve this, we use Berry's transitionless quantum driving algorithm which allows us to time evolve eigenstates suppressing transitions between them \cite{berry_transitionless_2009}.  We implement counterdiabatic driving in two configurations as we describe in further detail below. In the first scheme, which we call the SC1 (shortcut 1) scheme, we add the counterdiabatic driving term directly to our Hamiltonian, implementing direct connections between the ground states as shown in Fig. \ref{fig1}(a). This is a situation suitable when it is possible to directly couple the lower energy states (e.g. using radio frequency transitions between ground states).  In the second scheme, which we call the SC2 (shortcut 2) scheme, we implement counterdiabatic driving by including them as Raman transitions, and thereby incorporating them into the existing SQR pulses as shown in Fig. \ref{fig1}(b).  This results in modified pulses that implement both the STIRAP and counterdiabatic driving term together, in a similar fashion to that performed in Ref. \cite{du_2016}.

\subsubsection{Counterdiabatic driving with direct transitions (SC1)}

Our starting point for determining the counterdiabatic driving is (\ref{kisrenzoniham}). As discussed in Ref. \cite{berry_transitionless_2009}, the counterdiabatic driving  corresponding can be obtained using the relation
\begin{align}
    H_{\text{cd}} & = i\hbar \sum_{m \neq n} \sum_{n=0}^3 \frac{\vert E_m \rangle \langle E_m \vert \partial_t H_{\text{SQR}}\vert E_n \rangle \langle E_n \vert}{E_n - E_m}  \label{hcd hsqr} \\
    & = i ( \omega_1 \vert 2 \rangle \langle 1 \vert  + \omega_2 \vert 3 \rangle \langle 1 \vert   +  \omega_3 \vert 3 \rangle \langle 2 \vert ) + \text{H.c.} ,
    \label{omegaham}
\end{align} 
where the sum over $n,m$ runs over nondegenerate eigenstates $\vert E_n \rangle$. Evaluating (\ref{hcd hsqr}) gives a $ 4 \times 4 $ Hamiltonian which can be written as a completely off-diagonal matrix, with purely imaginary elements such that $ \omega_i $ are real numbers. 

Our shortcut Hamiltonian then consists of combining this with (\ref{kisrenzoniham}) according to
\begin{equation}
    H_{\text{SC1}} = H_{\text{cd}} + H_{\text{SQR}}. \label{htotalfig1a}
\end{equation} 
This gives us the shortcut Hamiltonian corresponding to the implementation depicted in Fig. \ref{fig1}(a).

\subsubsection{Counterdiabatic driving via Raman transitions (SC2)}

We now move onto the second implementation, where the counterdiabatic driving terms are implemented by Raman transitions.  We begin by writing an 5-level ansatz Hamiltonian corresponding to Fig. \ref{fig1}(b). The Hamiltonian is given by 
\begin{equation}
    H_{\text{SC2}} = -\begin{pmatrix}
        \delta_1 & 0 & 0 & \tilde{\Omega}_1 & 0 \\ 0 & \delta_2 & 0 & \tilde{\Omega}_2 & \tilde{\Omega}_4 \\ 0 & 0 & \delta_3 & \tilde{\Omega}_3 & \tilde{\Omega}_5 \\ \tilde{\Omega}_1 & \tilde{\Omega}_2 & \tilde{\Omega}_3 & \Delta & 0 \\ 0 & \tilde{\Omega}_4 & \tilde{\Omega}_5 & 0 & \Delta 
    \end{pmatrix} , \label{H5}
\end{equation} 
where the first three levels are the same ground states as in the original SQR scheme, the state $ \vert  4 \rangle $ is the same excited state as in the SQR scheme, and $ \vert  5 \rangle $ is an excited state that is necessary to implement the counterdiabatic driving. The detuning $ \Delta $ is the same as that appearing in (\ref{kisrenzoniham}).  Note that we assume that $ \Delta >0 $ in (\ref{H5}).  $\delta_i$ and $\tilde{\Omega}_j$ are the parameters that are to be determined.  

The Hamiltonian (\ref{H5}) must implement both the SQR pulses and the counterdiabatic driving term. 
In the approach of Ref. \cite{du_2016}, the effective Hamiltonians of the STIRAP pulses in the adiabatically eliminated space (i.e. the space only containing ground states) are first derived. The counterdiabatic driving terms are calculated based on this effective Hamiltonian.   One obtains the total effective transitions by adding these contributions.  In our case, this is written
\begin{equation}
    H_{\text{SC2}}^{\text{eff}} = H_{\text{cd}}^{\text{eff}} + H_{\text{SQR}}^{\text{eff}} ,\label{criteria}
\end{equation} 
where we have put the label ``eff'' when we work in the adiabatically eliminated space.  
Finally, the Hamiltonian in the larger space including excited states is deduced from the combined Hamiltonian.  

We now evaluate each of the terms in (\ref{criteria}) explicitly to obtain a system of equations that will solve for the parameters in (\ref{H5}).  First, the adiabatically eliminated Hamiltonian corresponding to (\ref{kisrenzoniham}) is given by \cite{Jerke2007}
\begin{equation}
    H_{\text{SQR}}^{\text{eff}} = \frac{1}{4\Delta} \begin{pmatrix}
    \vert \Omega_1 \vert ^2 & \Omega_2\Omega_1^* & \Omega_3\Omega_1^* \\ \Omega_1\Omega_2^* & \vert  \Omega_2 \vert ^2 & \Omega_3\Omega_2^* \\ \Omega_1\Omega_3^* & \Omega_2\Omega_3^* & \vert  \Omega_3 \vert ^2 \\
\end{pmatrix}. \label{Hsqreff}
\end{equation}
The counterdiabatic driving terms corresponding to this Hamiltonian are defined by
\begin{align}
    H_{\text{cd}}^{\text{eff}} = i\hbar \sum_{m \neq n} \sum_{n=0}^2 \frac{\vert E_m \rangle \langle E_m \vert \partial_t H_{\text{SQR}}^{\text{eff}} \vert E_n \rangle \langle E_n \vert}{E_n - E_m} ,
    \label{hcdeff}
\end{align}
where $\vert E_n \rangle$ are the eigenstates and $E_n $ 
are the eigenenergies of (\ref{Hsqreff}).  Finally, the adiabatically reduced version of (\ref{H5}) can be written as
\begin{align}
    \begin{aligned}
    H_{\text{SC2}}^\text{eff}  = \frac{1}{\Delta} & \begin{pmatrix}
    \vert \tilde{\Omega}_1\vert^2  & \tilde{\Omega}_2\tilde{\Omega}_1^* & \tilde{\Omega}_3\tilde{\Omega}_1^* \\ \tilde{\Omega}_1\tilde{\Omega}_2^* & \vert \tilde{\Omega}_2\vert^2 + \vert \tilde{\Omega}_4\vert^2  & \tilde{\Omega}_2\tilde{\Omega}_3^* + \tilde{\Omega}_5\tilde{\Omega}_4^* \\ \tilde{\Omega}_1\tilde{\Omega}_3^* & \tilde{\Omega}_3\tilde{\Omega}_2^* + \tilde{\Omega}_4\tilde{\Omega}_5^* & \vert \tilde{\Omega}_3\vert^2 + \vert \tilde{\Omega}_5 \vert^2  \\
\end{pmatrix}  \\
& - \begin{pmatrix}
    \delta_1 & 0 & 0 \\ 
  0  &  \delta_2 & 0 \\ 
  0 & 0 & \delta_3 \\
\end{pmatrix}.  
    \end{aligned}\label{h5eff}
\end{align}

Let us begin by solving for the detunings $\delta_j$. Examining the diagonal elements of (\ref{criteria}), we have 
\begin{align}
\delta_1 & = \frac{4\vert \tilde{\Omega}_1\vert^2 - \vert \Omega_1\vert^2}{4\Delta} \\
\delta_2 & = \frac{(4(\vert \tilde{\Omega}_2\vert^2 + \vert \tilde{\Omega}_4\vert^2) - \vert \Omega_2\vert^2)}{4\Delta} \\
\delta_3 & = \frac{(4(\vert \tilde{\Omega}_5\vert^2 + \vert \tilde{\Omega}_3\vert^2) - \vert \Omega_3\vert^2)}{4\Delta}.
\end{align}
There is no contribution of $ H_{\text{cd}}^{\text{eff}}$ since it is a completely off-diagonal matrix. 

Next, we solve for the amplitudes $ \tilde{\Omega}_i $. Examining the off-diagonal elements of (\ref{criteria}), we obtain a system of three equations, not counting Hermitian conjugates. For the $ i$th row and $ j$th column we have
\begin{equation} 
\frac{\tilde{\Omega}_i^*\tilde{\Omega}_j}{\Delta}  = \frac{\Omega_i \Omega_j^*}{4\Delta} + \langle i \vert H^{\text{eff}}_{\text{cd}} \vert j \rangle,  \label{effectiveterm}
\end{equation}
For convenience of subsequent expressions let us define the quantity
\begin{equation}
R_{ij}e^{i\varphi_{ij}} \equiv \frac{\Omega_i \Omega_j^*}{4} + \Delta \langle i \vert H^{\text{eff}}_{\text{cd}} \vert j \rangle,
\label{relation} 
\end{equation}
which is the right hand side of (\ref{effectiveterm}) multiplied by $ \Delta $.  Note that $R_{ij}$ and $\varphi_{ij}$ are known quantities, as we have the explicit form of the SQR pulses $ \Omega_i $, and the counterdiabatic driving Hamiltonian is easily calculated by diagonalizing a $ 3 \times 3 $ Hamiltonian (\ref{hcdeff}).  We do not write explicit expressions as they are cumbersome and are better handled numerically.  

We may now solve for the amplitudes $ \tilde{\Omega}_i $.  From (\ref{relation}) the off-diagonal elements are
\begin{align} 
\tilde{\Omega}_2\tilde{\Omega}_1^*    & =  R_{12}e^{i\varphi_{12}} \label{polardecomp12} \\
  \tilde{\Omega}_3\tilde{\Omega}_1^*  & =  R_{13}e^{i\varphi_{13}}  \label{polardecomp13} \\
  \tilde{\Omega}_2\tilde{\Omega}_3^* + \tilde{\Omega}_5\tilde{\Omega}_4^*  & = R_{23}e^{i\varphi_{23}}  . \label{polardecomp}
\end{align} 
We first solve for $\tilde{\Omega}_1$ and $\tilde{\Omega}_2$. Eq. (\ref{polardecomp12}) is under-constrained with respect to the variables $\tilde{\Omega}_1$ and $\tilde{\Omega}_2$, hence we may choose a convenient solution.  We divide the amplitude and phase equally between the two variables, and choose the solution
\begin{align}
\tilde{\Omega}_1 & =  e^{-i\frac{\varphi_{12}}{2}} \sqrt{R_{12}}  \label{omega1def} \\
\tilde{\Omega}_2 & = e^{i\frac{\varphi_{12}}{2}} \sqrt{R_{12}}
\end{align} 

Next, we solve for $\tilde{\Omega}_3$ by substituting (\ref{omega1def}) into (\ref{polardecomp13}). Doing so gives us 
\begin{equation}
    \tilde{\Omega}_3 = \frac{R_{13}}{\sqrt{R_{12}}} e^{i(\varphi_{13} + \frac{\varphi_{12}}{2})}. \label{omega3def}
\end{equation} 
Due to the presence of $ R_{12} $ on the denominator, this expression can have some undesirable features when $ R_{12} $ is small.  Such features can be handled since they occur in regions where all pulses are small.  In Appendix \ref{appendix sec 1} we show methods of removing such features with a modulation envelope which has minimal effect on the evolution of the pulses. 

Finally, we solve for $\tilde{\Omega}_{4}$ and $\tilde{\Omega}_{5}$. Again using the fact that $ \tilde{\Omega}_2 $ and $ \tilde{\Omega}_3 $ are already solved for, we rearrange (\ref{polardecomp}) and define
\begin{equation}
 R e^{i \varphi } \equiv R_{23}e^{i\varphi_{23}} - \tilde{\Omega}_2\tilde{\Omega}_3^*. \label{newelement23}
\end{equation} 
This is again an under-constrained equation and we choose the solution
\begin{align}
 \tilde{\Omega}_{4} & = e^{-i\frac{\varphi}{2}} \sqrt{R} \\
 \tilde{\Omega}_{5} & = e^{i\frac{\varphi}{2}} \sqrt{R} 
\end{align}
This concludes the determination of all parameters of (\ref{H5}).

In Ref. \cite{du_2016} the counterdiabatic driving on a 3-level $\Lambda $ system could be implemented with modified pulses on a system of the same dimensionality. Here, an additional level was introduced (for a total of 5 levels) to implement the SQR pulses and the counterdiabatic driving, in comparison to the original 4-level $\Lambda$ SQR scheme. The reader may wonder whether it is possible to reduce the number of levels in Fig. 1(b) to four levels, further simplifying the protocol.  We show in Appendix B that this is not possible in the context of STIRAP quantum gates, due to an insufficient number of degrees of freedom in the 4-level scheme.  This is a fundamental difference to the results of Ref. \cite{du_2016} where the counterdiabatic terms can  be incorporated into a modified STIRAP pulse without introducing another level.

The reader may also wonder why the counterdiabatic driving terms must be added in the adiabatically eliminated space (\ref{criteria}).  This is a simpler procedure than adding them in the full space including the excited states because of the combination of two Raman pulses does not generally lead to sum of the effective couplings.  For example, suppose we wish to add the effect of two Raman couplings between levels $ |1 \rangle$ and $ | 2 \rangle $, where the two effective couplings are $ \Omega_1 \Omega_2^*/\Delta $ and $\Omega_1' \Omega_2'^*/\Delta $.  Simply adding the transitions together produces the overall coupling $ (\Omega_1 + \Omega_1')(\Omega_2^* + \Omega_2'^*)/\Delta \ne  \Omega_1 \Omega_2^*/\Delta + \Omega_1' \Omega_2'^*/\Delta  $.  As such, first working in the effective space and then deducing the pulses required to produce the combined coupling is simpler procedure to determine the modified pulses.

\begin{figure*}[t]
\includegraphics[width=\linewidth]{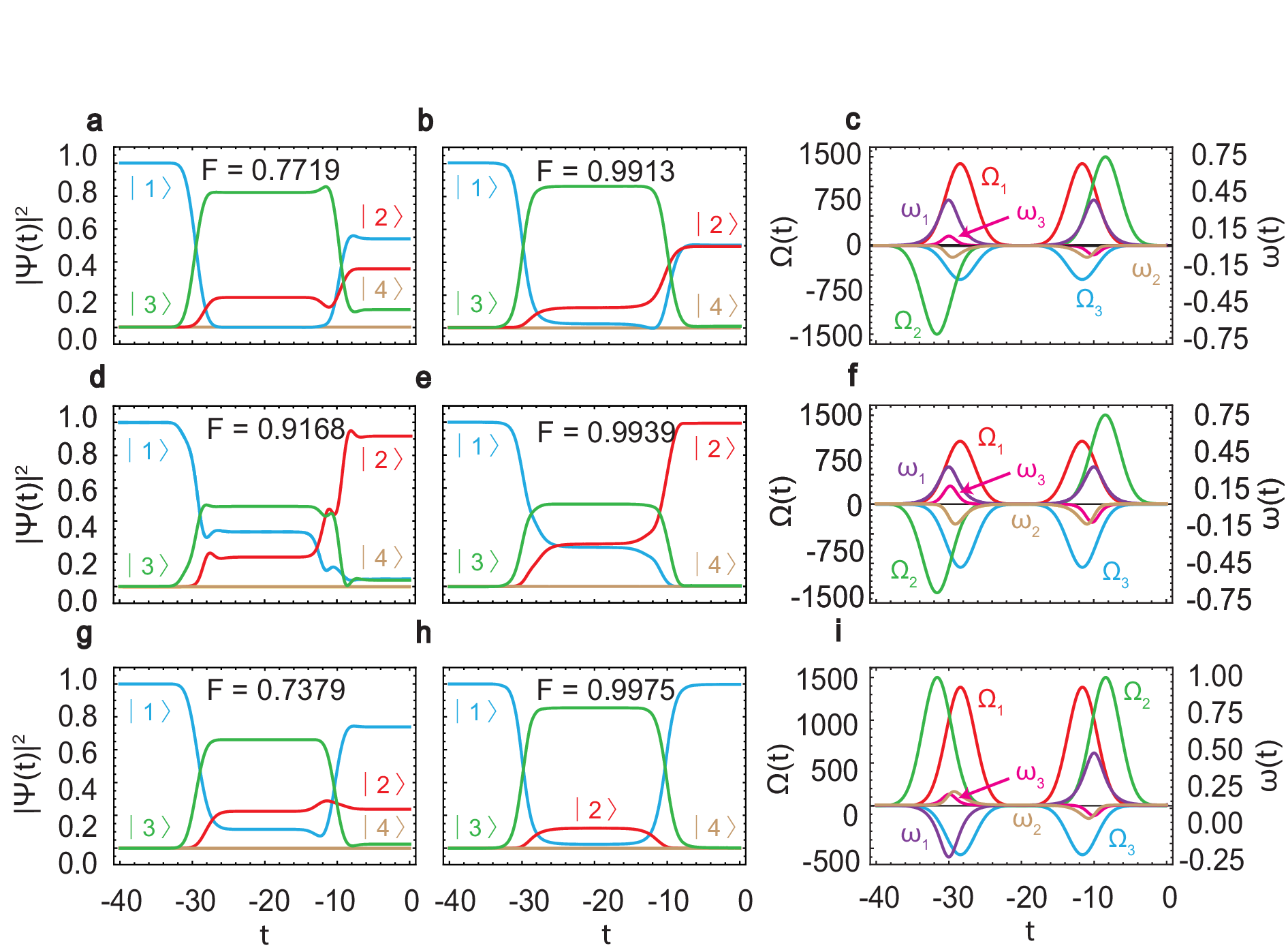}
\setlength{\abovecaptionskip}{-0.25cm}
\caption{Performance of counterdiabatic driving with direct transitions (scheme SC1).  The top row ((a)(b)(c)) corresponds to the Hadamard gate, the second row ((d)(e)(f)) a Pauli-$X$ gate, the bottom row ((g)(h)(i)) the identity gate.  In the left column ((a)(d)(g)) we show the evolution of the bare SQR scheme without counterdiabatic driving, using Hamiltonian (\ref{kisrenzoniham}).  In the middle column ((b)(e)(h))  the counterdiabatic driving is included according to Hamiltonian (\ref{htotalfig1a}). The fidelities (\ref{fiddef})  are marked in each time evolution plot. The right column shows the pulses that drive the rotation, where $ \Omega_i $ are the SQR pulses and $ \omega_i $ are the counterdiabatic driving pulses.  The parameters for each gate are $\Omega_0 = 250$, $\chi = \frac{\pi}{8}$, $\eta = \pi$, and $\delta = 0$ for the identity gate; $\Omega_0 = 750$, $\chi = \frac{\pi}{4}$, $\eta = \pi$, and $\delta = \pi$ for a Pauli-$X$ gate; and $\Omega_0 = 350$, $\chi = \frac{\pi}{8}$, $\eta = \pi$, and $\delta = \pi$ for a Hadamard gate. For all plots, values of $T = 20$, $t_0 = 1.6$, and $\sigma = 2.0$ initial condition $\vert \psi \rangle = \vert 1 \rangle$ were used.
\label{fig3}  }
\end{figure*}

\section{Numerical Simulations}
\label{sec:numerics}
\subsection{Simulation Methods}
\label{iii a: simulation methods}

To analyze the efficacy of our methods, simulations were performed for the SQR scheme with and without the counterdiabatic driving.  To do so, we time evolve the quantum state with respect to either the Hamiltonian (\ref{htotalfig1a}) for scheme SC1 or the Hamiltonian (\ref{H5}) for scheme SC2 by the time dependent Schrodinger equation 
\begin{equation}
    i\hbar \frac{d}{dt}\vert \psi \rangle = H_{\text{SC}} (t) \vert \psi \rangle.
    \label{schrodingereqn}
\end{equation} 
All states are evolved from initial state $\vert \psi \rangle = \vert 1 \rangle$. We solve the Schrodinger equation numerically in Mathematica and plot the population of each state across the entire time scale. The fidelity is evaluated by calculating 
\begin{align}
F= \vert  \langle \psi(t_{\max}) \vert   U_{\hat{n}}(\delta)\vert \psi_0 \rangle \vert ^2  ,
\label{fiddef}
\end{align}
where $ \vert   \psi(t_{\max}) \rangle $ is the state at the end of the evolution (\ref{schrodingereqn}) and the comparison state is (\ref{rotationformula}).  

We demonstrate our protocol on three quantum gates: the Hadamard gate, the Pauli-$X$ gate, and the identity gate.  We expect that the shortcut Hamiltonians have a performance improvement in terms of an increased fidelity in the  diabatic regime. The most suitable regime of interest is near the boundary of the adiabaticity.  We therefore choose the STIRAP pulse parameters (\ref{pulse1})-(\ref{pulse3}) to be initially of poor fidelity, then incorporate the counterdiabatic driving to see whether the fidelity can be improved.  We note that increasing the overall amplitude $ \Omega_0 $ of the pulses is equivalent to making the timescale longer.  For example, in the unitary evolution $ e^{-iHt/ \hbar }$, to achieve the same effect as a longer evolution $ t \rightarrow at $ with $ a> 1$, we could equally multiply the Hamiltonian by a factor $ H \rightarrow a H $. By choosing a larger $ \Omega_0 $, we may therefore push the system in and out of the adiabatic regime according to our choosing.

\subsection{Counterdiabatic driving with direct transitions (SC1) \label{iii b: SQR with direct cd driving simulations}} 

We first examine the performance of the shortcut Hamiltonian (\ref{htotalfig1a}) which implements the counterdiabatic driving using direct connections between the ground states (scheme SC1, Fig. \ref{fig1}(a)).  Figure \ref{fig3} shows our numerical results for the three quantum gates in each row.  In the left column (Fig. \ref{fig3}(a)(d)(g)), we  show the performance of the standard SQR scheme by time evolving (\ref{kisrenzoniham}), choosing parameters in the regime where adiabadicity is starting to break down.  In the middle column we implement the counterdiabatic driving (Fig. \ref{fig3}(b)(e)(h)).  In the right column, we show the pulses that are required to implement the scheme (Fig. 3(c)(f)(i)).  The fidelities (\ref{fiddef}) of the original SQR scheme and the shortcut scheme are shown on each plot. 

We see that for all gates the counterdiabatic driving term is effective in improving population transfer corresponding to the qubit rotation. Fidelities are also improved for all cases.  However, the degree to which the performance is enhanced varies across gates. We have chosen parameters such that the fidelities of the gates with counterdiabatic driving are approximately at the $ 99 \% $ level.  As can be seen from Fig. \ref{fig3}(d), this corresponds to an original fidelity of the bare SQR protocol to be $ \sim 92 \% $, which is much higher than the other two gates.  As such, we see the best improvement in the identity gates and the weakest performance improvement in the Pauli-$X$ gate. While we do not have a full explanation of the performance enhancement, we note that the trend appears to be that the more orthogonal the ideal final state is to the initial state, the less the performance improvement.  

Examining the evolution curves for the standard SQR protocol in the diabatic regime (Fig. \ref{fig3}(a)(d)(g)), we see that in all cases there is a non-negligible population in the auxiliary ground state $\vert 3 \rangle$.  This represents incomplete state transfer back to the basis that defines the qubit ${\vert 1 \rangle, \vert 2 \rangle}$. This is unideal as it represents an incomplete gate. The addition of counterdiabatic driving serves to remove this remaining population of the auxiliary state (Fig. \ref{fig3}(b)(e)(h)), which shows that the shortcut Hamiltonian is effective at suppressing such imperfections. 
We note that amplitudes of the correction pulses $\omega_i $ are small in comparison to the SQR pulses $\Omega_i$ (Fig. \ref{fig3}(c)(f)(i)). This shows that we are not solely increasing the magnitude of the Raman pulses (which pushes the system into the adiabatic regime) but are adjusting the control in a nontrivial manner. We also point out that the phase relations of the pulses are simple in this case. The SQR pulses $\Omega_i$, and the counterdiabatic driving pulses are purely imaginary. This simplifies the implementation of the counterdiabatic driving term for the SC1 scheme, whereas we will see a more complex phase relation for the SC2 scheme.

\begin{figure}[t]
\includegraphics[width=\linewidth]{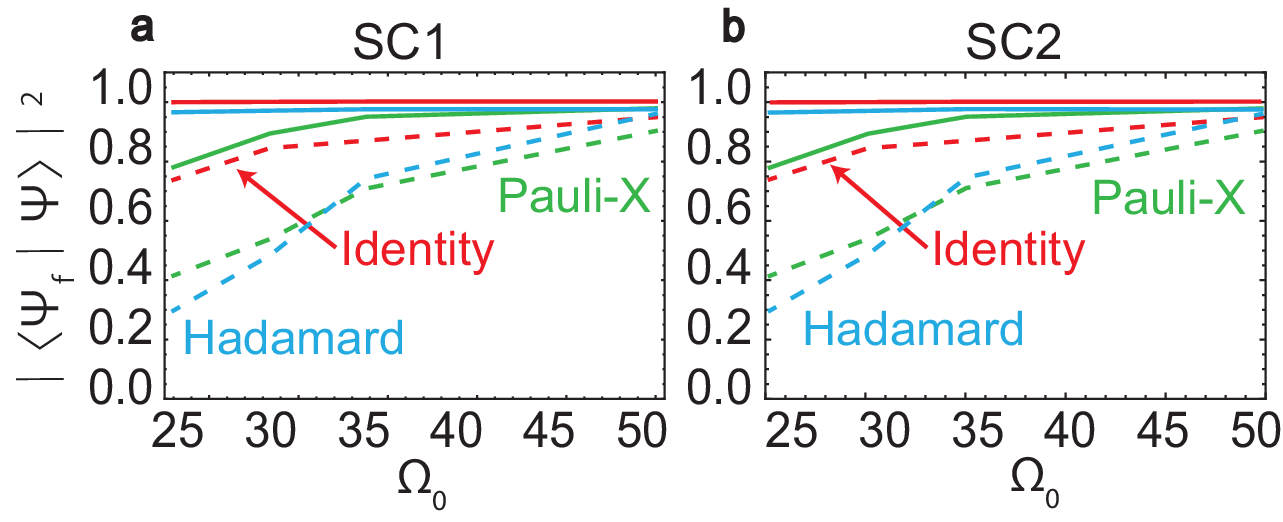}
\caption{Fidelity dependence on pulse amplitude (solid lines) for the shortcut Hamiltonians (a) Eq. (\ref{htotalfig1a}), SC1 scheme; (b) Eq. (\ref{H5}), SC2 scheme.  For comparison on both plots we show the equivalent fidelities (dashed lines) without counterdiabatic driving using Hamiltonian (\ref{kisrenzoniham}). All time evolutions were performed using the parameters $\Delta = 10\Omega_0, T = 20, t_0 = 1.6,$, and $\sigma = 2$. 
\label{fig5}  }
\end{figure}

Fig. \ref{fig5}(a) shows the fidelity for a range of amplitudes for the three gates as a function of the overall amplitude $ \Omega_0 $.  As discussed above, a larger amplitude puts the system further into the adiabatic regime, resulting in an improvement of all the gate fidelities.  The counterdiabatic driving allows for smaller amplitudes (or equivalently shorter STIRAP pulses) to be used, while maintaining a similar fidelity.  The Pauli-$X$ gate experiences the earliest reduction in fidelity as the pulse amplitudes are decreased, showing that it is the most susceptible to diabatic excitations.

\begin{figure*}[t]
\includegraphics[width=\linewidth]{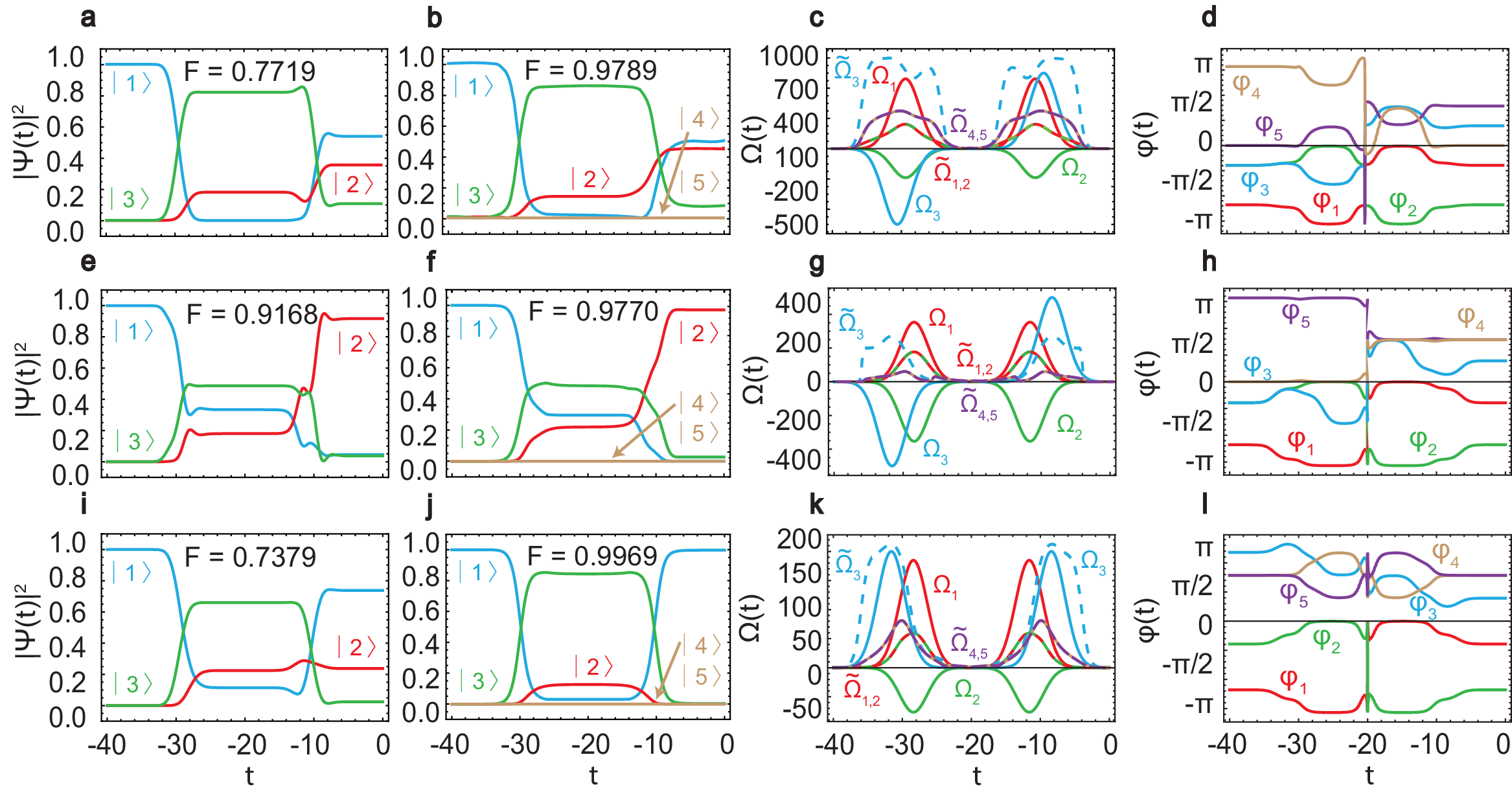}
\setlength{\abovecaptionskip}{-0.0cm}
\caption{Performance of counterdiabatic driving via Raman transitions (scheme SC2).  The first row (a)-(d) corresponds to the Hadamard gate, the second row (e)-(h) the Pauli-$X$ gate, the third row (i)-(l) the identity gate. The left column (a)(e)(i) shows the evolution of the bare SQR scheme without counterdiabatic driving, using Hamiltonian (\ref{kisrenzoniham}). Column (b)(f)(j) shows the time evolution including counterdiabatic driving according to Hamiltonian (\ref{H5}). Fidelities for each are computed according to (\ref{fiddef}) and marked on each figure. Column (c)(g)(k) shows the amplitudes of the pulses of the SQR Hamiltonian (\ref{kisrenzoniham}) (dashed lines), and the SC2 Hamiltonian (\ref{H5}) (solid lines).  The column (d)(h)(l) shows the phases of the pulses in the SC2 Hamiltonian (\ref{H5}). 
 All simulations are run with the initial state $\vert \psi \rangle = \vert 1 \rangle$ and parameter set $\Omega_0 = 500$, $\Delta = 100\Omega_0$, $T = 20.0$, $t_{\text{min}} = -2T$, $t_{\text{max}} = 0$, $\sigma = 2$, and $t_0 = 1.6$. For each row, the rotation parameters $\chi$, $\eta$, and $\delta$ are the same as in Fig. \ref{fig3}.  
\label{fig4}  }
\end{figure*}

\subsection{Counterdiabatic driving via Raman transitions (SC2)}
\label{iii c: SQR with lambda cd driving simulations}

We now examine the performance of the shortcut Hamiltonian (\ref{H5}), which implements the counterdiabatic driving using Raman transitions as discussed in Sec. \ref{iii b: SQR with direct cd driving simulations} (scheme SC2, Fig. \ref{fig1}(b)).  Figure \ref{fig4} show the results of our numerical simulations for the same quantum gates as before in each row. The left column (Fig. \ref{fig4}(a)(e)(i)) shows performance of the standard SQR scheme time evolving (\ref{kisrenzoniham}), choosing parameters in the regime where adiabadicity is starting to break down.  We then implement the counterdiabatic driving according to (\ref{H5}) as shown in the middle left column (Fig. \ref{fig4}(b)(f)(j)). The fidelities of the original SQR scheme and the improved scheme are shown on each plot. Finally, we show the pulses required to implement the SQR scheme in the right two columns, showing the magnitude (Fig. \ref{fig4}(c)(g)(k)) and phases (Fig. \ref{fig4}(d)(h)(l)). 

We see that in all cases the modified pulses including the counterdiabatic driving improves the final fidelity of the gates.  
The initial SQR pulses are in a diabatic regime, where the states begin in the qubit state space ${\vert 1 \rangle, \vert 2 \rangle}$, but do not make a full return to this subspace upon being rotated. This is improved by the addition of the counterdiabatic driving term, where the population of the state $ \vert  3 \rangle $ is lower than the bare SQR Hamiltonian (Fig. \ref{fig4}(b)(f)(j)).  We note that there is still some population in the state $ \vert  3 \rangle $ for the Hadamard gate (Fig. \ref{fig4}(b)). In comparison, for scheme SC1, this population was nearly completely eliminated (Fig. \ref{fig3}(b)).  Nevertheless, there is a considerable fidelity improvement, albeit not quite as high as in scheme SC1.  We attribute the slight performance decrease to the fact that the implementation of the shortcut scheme SC2 requires approximate expressions (\ref{Hsqreff}) and (\ref{h5eff}) working in the adiabatically eliminated space.  The excited states $\vert 4 \rangle$ and $\vert 5 \rangle$ however have low populations throughout.

The pulse amplitudes of the shortcut Hamiltonian as shown by the dashed lines in Fig. \ref{fig4}(c)(g)(k), are comparable to the original STIRAP pulses $\Omega_i$. Here, the amplitudes $ \tilde{\Omega}_1,\tilde{\Omega}_2,\tilde{\Omega}_3 $ implement both the SQR pulses as well as the counterdiabatic driving. Due to the combination of the two contributions we see that the pulses are no longer a Gaussian shape, which is consistent with Ref. \cite{du_2016} where a similar procedure was performed for the canonical 
3-level STIRAP.  Amplitudes $ \tilde{\Omega}_4,\tilde{\Omega}_5$ purely implement corrections to the  counterdiabatic driving alone between levels $ \vert 2 \rangle$ and $ \vert  3 \rangle $.  The pulses $ \tilde{\Omega}_4,\tilde{\Omega}_5$ are again non-Gaussian because they are corrections to produce the counterdiabatic driving.  As shown in (\ref{newelement23}), this transition implements the difference to the $ \Lambda $ transition through pulses $ \tilde{\Omega}_2, \tilde{\Omega}_3 $.  Since $ \tilde{\Omega}_2 , \tilde{\Omega}_3 $ are non-Gaussian, the difference implemented by $ \tilde{\Omega}_4,\tilde{\Omega}_5$ is also non-Gaussian.  

The phases of the modified pulses give a more complex relation, as can be seen in Fig. \ref{fig4}(d)(h)(l).  This is due to the combination of the completely real SQR pulses with the purely imaginary counterdiabatic driving terms.  A similar situation was found in Ref. \cite{du_2016}, but the phase could be eliminated by a suitable unitary rotation. There is no such unitary rotation that would eliminate the phases in our scheme. This is due to the interdependence of the effective terms. In the case of Ref. \cite{du_2016}, there was one term that captures the effective connection of the system. In our case, we have three terms that form an interdependent system (see Appendix \ref{appendix sec 1} for further discussion of this point). Performing a unitary transformation to eliminate one phase will inevitably influence the other phases. We add here that the phase discontinuity at the middle of the two STIRAP sequences may appear problematic in a real implementation.  However, examining the amplitudes Fig. \ref{fig4}(c)(g)(k) reveals that the discontinuities occur when the pulse amplitudes are very small. Consider the first and second set of pulses as separate pulses, the discontinuities should be relatively harmless.

Figure \ref{fig5}(b) shows the dependence of the fidelity for the three gates as a function of the pulse amplitude $ \Omega_0$, which controls the adiabaticity.  Similarly to the SC1 schematic, larger pulse amplitudes push the system into an adiabatic regime, resulting in enhanced fidelity in all cases. The degree of improvement remains gate dependent. The Pauli-$X$ gate again experiences earliest reduction in fidelity, while the Hadamard gate experiences the largest degree of improvement. The similarity between the fidelity improvement of SC1 and SC2 implementations is encouraging. The fidelities differ only by magnitudes of $10^{-2}$, which can be attributed to the approximations required in implementing SC2, as mentioned in the previous section. This suggests that both implementations are equally amenable to enhancing qubit rotation.

We note that we only considered three quantum gates and an initial state set to $ \vert  \psi_0 \rangle = \vert  1 \rangle $.  We have tested our code for several other gates and initial states and have found similar performance to what we have shown in Figs. \ref{fig3} and \ref{fig4}, hence we consider these representative of what is obtained for other gates and initial states.

\section{Summary and conclusions \label{conclusion}}

In this paper, we have developed a shortcut to adiabaticity approach to improve the performance of the SQR protocol introduced in Ref. \cite{Kis2002}. The basic idea is to apply the counterdiabatic driving as given in Ref. \cite{berry_transitionless_2009}, which acts to suppress unwanted excitations into higher eigenstates. To implement this method, we presented two ways of implementing the shortcut Hamiltonian to improve gate performance. In the SC1 scheme, the counterdiabatic driving was introduced directly between the ground states. This corresponds to systems where it is possible to address the individual ground states, perhaps using radio frequency or microwave pulses. In the SC2 scheme, we introduce Raman transitions to implement counterdiabatic driving on a 5-level $\Lambda$ scheme.  This allows us to merge the counterdiabatic driving with the original SQR pulses, resulting in modified pulses with a non-Gaussian form and special phase dependence. The form of the pulses are easily calculated numerically, although they do not have a simple analytical form. We then numerically simulated the time evolution of these Hamiltonians to demonstrate their efficacy in performing quantum gates. 

Our results demonstrate that it is possible to improve the SQR scheme  using shortcut to adiabaticity. We demonstrated the enhanced performance of both the SC1 and SC2 schemes by comparing the fidelity of the final state for three quantum gates. As expected, the shortcut methods are most effective in the regime where adiabaticity breaks down (Fig. \ref{fig5}).  The counterdiabatic driving is helpful to restore adiabaticity, and can attain high fidelity gates in all cases considered.  The methods have their limits, and if the original SQR Hamiltonian is too deep in the diabatic regime, the counterdiabatic terms cannot recover high fidelity gate operations.  The price to be paid for this are either additional controls between the ground states for the SC1 scheme, or modified pulse amplitudes and phases as in the SC2 scheme.  Further improvement upon the pulses may be able to be obtained via inverse engineering methods \cite{yan_experimental_2021}. 

This method is amenable to extension as long as the number of degrees of freedom is sufficient to solve the system of equations produced by the criteria (\ref{criteria}). As discussed in the Appendix \ref{appendix2}, using only a 4-level scheme produces an insufficient number of degrees of freedom to solve for modified pulses. In addition to the 5-level scheme that we have discussed in this paper (Fig. \ref{fig1}(b)), we have also shown that 6-level and 7-level schemes all allow for the implementation of counterdiabatic driving. The additional levels are used to implement the three $ \omega_i $ couplings as shown in Fig. \ref{fig1}(a).  We opted to discuss the 5-level configuration in this paper since this appeared to be the minimal system that could still implement the SQR pulses and counterdiabatic driving together.  The penalty for this is that the phases and amplitude dependence of the pulses are somewhat complex, as seen in Fig. \ref{fig4}.  Using a larger number of levels simplifies this as each coupling is no longer dependent on the others. Thus, this method of implementing counterdiabatic driving to achieve shortcut to adiabaticity is amenable to a variety of contexts performing quantum gates such as qudit rotations and entangling quantum gates.

\section{Acknowledgments}
We thank Dries Sels for discussions. This work is supported by the National Natural Science Foundation of China (62071301); NYU-ECNU Institute of Physics at NYU Shanghai; Shanghai Frontiers Science Center of Artificial Intelligence and Deep Learning; the Joint Physics Research Institute Challenge Grant; the Science and Technology Commission of Shanghai Municipality (19XD1423000,22ZR1444600); the NYU Shanghai Boost Fund; the China Foreign Experts Program (G2021013002L); the NYU Shanghai Major-Grants Seed Fund; Tamkeen under the NYU Abu Dhabi Research Institute grant CG008; and the SMEC Scientific Research Innovation Project (2023ZKZD55).

\begin{figure}[t]
\includegraphics[width=\linewidth]{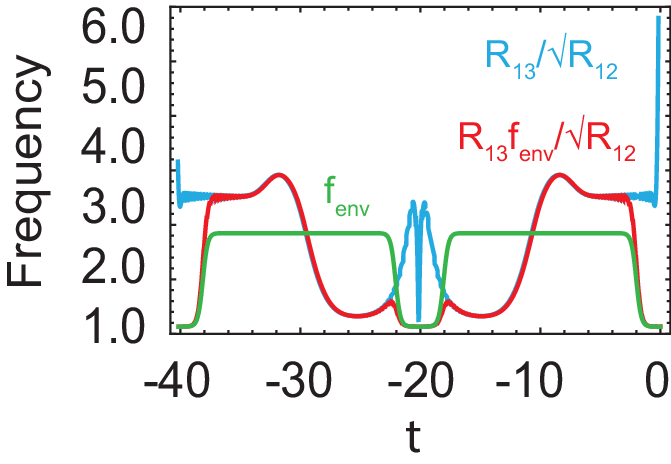}
\setlength{\abovecaptionskip}{-0.5cm}
\caption{Adjustment of pulse amplitude $\tilde{\Omega}_3$  using a modulation envelope function. The graph shows $f_{\text{env}}$ constructed to smoothen the pulse amplitude. The effect of the application of $f_{\text{env}}$ can be seen by comparing the  amplitude $\frac{R_{13}}{\sqrt{R_{12}}}$ and the amplitude $\frac{R_{13}f_{\text{env}}}{\sqrt{R_{12}}}$. 
\label{fig6}  }
\end{figure}

\appendix
\section{The expression for  $\tilde{\Omega}_3$ in Eq. (\ref{omega3def}) }
\label{appendix sec 1}

In solving for the modified Rabi pulse $\tilde{\Omega}_3$ in (\ref{omega3def}) we encounter a term $\frac{R_{13}}{\sqrt{R_{12}}}$. This term can be problematic given the behavior of $R_{12}$ and $R_{13}$. Both magnitudes go to zero outside of the application of the pulses, but they differ in the rate at which they go to zero. This causes $\frac{R_{13}}{\sqrt{R_{12}}}$ to remain nonzero outside of the pulse application region, despite $R_{13} \rightarrow 0 $. An example of this is shown in Figure \ref{fig6}. We see that at and the start and end of the evolution $\frac{R_{13}}{\sqrt{R_{12}}}$ is non-zero and ill-behaved due to the numerically small values involved in the ratio.

To remedy this issue, we make use of a suitable modulation envelope function to send $R_{13}$ to zero outside of the pulse application regime. To do so, we define the function 
\begin{equation}
    f_{\text{env}}(t) = \frac{1 - \frac{1}{1 + e^{\Omega_0(t + \frac{T}{2} + 4\sigma)}}}{1 + e^{\Omega_0(t + \frac{T}{2} - 4\sigma)}} + \frac{1 - \frac{1}{1 + e^{\Omega_0(t + \frac{3T}{2} + 4\sigma)}}}{1 + e^{\Omega_0(t - \frac{3T}{2} + 4\sigma)}}.
\end{equation} 
The form of this function designed such that is equal to one in the region of pulse activity and zero outside of pulse activity region. 

We apply this function to the ratio $\frac{R_{13}}{\sqrt{R_{12}}}$ in our definition of $\tilde{\Omega}_3$ and use in our numerical routines the modified pulse
\begin{equation}
    \tilde{\Omega}_3  = \frac{R_{13}f_{\text{env}}}{\sqrt{R_{12}}}e^{(i\varphi_{13} - \frac{\varphi_{12}}{2})}.
\end{equation} 
Figure \ref{fig6} shows the effect of applying $f_{\text{env}}$. The application of this function forces the ratio of the magnitudes to be zero outside of the regions of pulse activity. The function does not influence activity within regions of pulse application since the function is set to one in those regions. With this modification, we are able to recover smooth behavior for $\tilde{\Omega}_3$.

\section{Counterdiabatic Driving with Raman transitions on a 4-Level $\Lambda$ System}
\label{appendix2}

Here we describe why for scheme SC2 we use a 5-level system as shown in Fig. \ref{fig1}(b), rather than the original 4-level system of the SQR Hamiltonian.  If one attempts to implement the counterdiabatic driving term on a 4-level system,  a lack of sufficient degrees of freedom to convert to a corresponding full Hamiltonian is encountered. Consider the off-diagonal elements of $H_{\text{SC2}}^{\text{eff}}$ as defined in (\ref{h5eff}). The values of these off-diagonal elements are set according to the right hand side of (\ref{criteria}) which in principle may be complete arbitrary.  However, the off-diagonal elements in (\ref{h5eff})  have an interdependence since they have the specific form $-\frac{\tilde{\Omega}_1^*\tilde{\Omega}_2}{\Delta}$, $-\frac{\tilde{\Omega}_1^*\tilde{\Omega}_3}{\Delta}$, $-\frac{\tilde{\Omega}_2^*\tilde{\Omega}_3}{\Delta}$.  Specifically, setting the phase of $ \tilde{\Omega}_i $ to be $ \phi_i$, then these three terms have a phase $ \xi_1 = \phi_2 - \phi_1$, $ \xi_2 = \phi_3 - \phi_1$, $ \xi_3 = \phi_3 - \phi_2$ respectively.  Since $ \xi_2 - \xi_1 = \xi_3 $, the phases of the off-diagonal elements cannot be chosen independently, and hence there is an insufficient number of degrees of freedom to satisfy  (\ref{criteria}). The addition of a 5th level to the scheme allows for an increase in degrees of freedom that can be used to write a proper corresponding Hamiltonian. 
\bibliography{paper}

\end{document}